\documentclass{article}

\usepackage{graphicx}
\usepackage{subfig}
\usepackage{epstopdf}
\usepackage{amsmath}
\usepackage{amsfonts}
\usepackage{authblk}
\usepackage{url}

\usepackage{natbib}
\newcommand{\DDELTA}{\texttt{Delta}\,}
\newcommand{\Eqnref}[1]{Eq.~\ref{eq:#1}}
\newcommand{\Eqsref}[1]{Eqs.~\ref{eq:#1}}
\newcommand{\Figref}[1]{Fig.~\ref{fig:#1}}
\newcommand{\Tabref}[1]{Tab.~\ref{tab:#1}}
\newcommand{\Secref}[1]{Sec.~\ref{sec:#1}}
\newcommand{\secref}[1]{sec.~\ref{sec:#1}}
\newcommand{\Te}{\ensuremath{T_\mathrm{e}}\,}

\title{Near real-time streaming analysis of big fusion data} 
\author[1]{R. Kube}
\author[1]{R.~M. Churchill}
\author[1]{CS Chang}
\author[2]{J. Choi}
\author[2]{J. Wang}
\author[2]{S. Klasky}
\author[3]{L. Stephey}
\author[4]{M.~J. Choi}
\author[5]{E. Dart}

\affil[1]{Princeton Plasma Physics Laboratory, Princeton, NJ 08540 USA}
\affil[2]{Oak Ridge National Laboratory, Oak Ridge, TN 37831 USA}
\affil[3]{Lawrence Berkeley National Laboratory, Berkeley, CA, USA}
\affil[4]{Korea Institute of Fusion Energy, Daejeon 34133, Republic of Korea}
\affil[5]{Energy Sciences Network}

\begin{document}

\maketitle


\begin{abstract}
While experiments on fusion plasmas produce high-dimensional data time series with ever increasing magnitude and velocity,
data analysis has been lagging behind this development. For example, many data analysis tasks are often performed in a
manual, ad-hoc manner some time after an experiment. In this article we introduce the \DDELTA framework that facilitates
near real-time streaming analysis of big and fast fusion data. By streaming measurement data from fusion experiments to
a high-performance compute center, \DDELTA allows to perform demanding data analysis tasks in between plasma pulses. 
This article describe the modular and expandable software architecture of \DDELTA and presents performance benchmarks of its 
individual components as well as of entire workflows. Our focus is on the streaming analysis of ECEi data measured at
KSTAR on NERSCs supercomputers and we routinely achieve data transfer rates of about $500$ MByte per second. We show that
a demanding turbulence analysis workload can be distributed among multiple GPUs and executes in under $5$ minutes. We further discuss
how \DDELTA uses modern database systems and container orchestration services to provide web-based real-time data visualization. 
For the case of ECEi data we demonstrate how data visualizations can be augmented with outputs from machine learning models. By providing
session leaders and physics operators results of higher order data analysis using live visualization they may monitor the
evolution of a long-pulse discharge in near real-time and may make more informed decision on how to configure the machine for the next shot.
\end{abstract}

\section{Introduction}
Experiments and plasma diagnostics of ever increasing sophistication are used to develop magnetic fusion solutions that will power
global energy demand. While current tokamaks operate in a pulsed manner, future fusion power plants need to operate in steady state
or with long pulses to deliver base load electricity. At the same time sample plasma diagnostics fusion plasma with ever increasing
spatial resolution and sampling frequencies, resulting in ever increasing data volumes and data rates from experiments.
\textbf{Get historic shot data sizes in MDS for NSTX}.
Under routine operation, ITER is expected to generate about two 
Petabytes of data per day. And while preliminary data analysis will take place on site, in-depth analysis of the
measurements as well as comparison with numerical simulations will take place at computational centers operated
by the individual ITER members.

Increasing data volume and data rates are met by a dichotomy of growth in compute power and growth in
capacity of the storage systems at High Perfomance Computing (HPC) systems. Over the last 20 years, the floating point operations
of all systems listed in the top500 list increased approximately tenfold every four years \footnote{\url{https://www.top500.org/statistics/perfdevel/}}.
The performance of the I/O systems of the Top500 supercomputer has only been tracked for the last four years,
and the peak bandwidth of the fastest system increased approximately four-fold. \footnote{\url{https://www.vi4io.org/io500/start}}.
With abundant computing resources and more scarce I/O resources, computationally intensive data analysis and workflows
that are based on that data, need to be implemented in a streaming manner in order to optimally scale with future
trends in HPC. That is, dropping the requirement to wait on the data to load from the I/O systems allows streaming
data analysis methods to make full use of advances in computing power. At the same time does streaming analysis offers streaming
analysis a faster turnaround time for data analysis.

In this paper we are presenting a streaming data analysis framework that aims to connect fusion experiments with
remote HPC resources \cite{choi-2016, kube-2020, kube-2020-ihpc, churchill-2021}. The streaming paradigm implemented by this
framework allows the big- and fast data generated by fusion experiments to be seamlessly analyzed using supercomputers as they are generated.
Our framework also includes a component for web-based near real-time data visualization. As soon as the data is analyzed it
can be visualized on a the visualization dashboard and at any place in the world connected to the internet.
By combining a robust streaming protocol with near real-time analytics and visualization we aim to tie together multiple threads
that are currently emerging in the analysis of fusion data. First, machine operators and session leaders may make more efficient
use of machine time by having access to higher order data analysis immediately after a plasma discharge has finished.
Second, interested scientists across the globe have direct access to results from experiments. And third, this approach provides
means to store large sets of analyized fusion data together with descriptive meta-data, making it suitable to continuously
train data-intensive machine learning models as experiments are commencing.

Current research in support of long-pulse steady state plasma operations focuses on the development of real-time
analysis algorithms for camera as well as thermographic data. At Wendelstein 7-X \cite{puig-stitjes-2018, jakubowski-2018, pisano-2020} 
and WEST \cite{mitteau-2021}, such developments aim to protect the machine from failures of actively cooled power exhaust solutions
as well as to monitor their performance. In order to do so, such systems analyze camera data and interact in real-time with a
local plasma control system. Using real-time control systems as downstream consumers of the data analysis sets stringent
limits on the runtime of the data analysis routines that can be performed on the camera data in this situation. 
At the other end of the spectrum in a fusion scientists toolbox lies integrated modeling. This describes the coupling of individual 
physics and engineering codes with the goal of modeling and understanding of plasma discharges \cite{poli-2018}. Data
analysis software, such as OMFIT \cite{OMFIT2015}, can be used to manage integrate individual codes into flexible
analysis workflows. As of now, these workflows are usually performed on local compute clusters and resort to
low-fidelity or time-independent models, when they are run in between pulses. Often, more demanding modelling is
run ad-hoc after an experiment has concluded.
Previous research work on automatic remote post-shot data analysis and simulation is described in \cite{kostuk-2018}.
Here, measurements from the DIII-D tokamak are used as input for numerical simulations that were performed on a
supercomputer at the Argonne National Laboratory Leadership Compute Facility immediately after the plasma pulse.
In this approach the supercomputer connected directly to the fusion experiment's MDSplus
\footnote{\url{http://www.mdsplus.org}} database and submitted simulations into a high-priority
queue using the supercomputers job scheduler.

In this article describe the design and implementation details of our a\textbf{D}aptive n\textbf{E}ar rea\textbf{L}-\textbf{T}ime \textbf{A}nalysis (\DDELTA)
framework. As a use-case we focus on the remote analysis of KSTAR electron cyclotron emission imaging (ECEi) measurements
using the Cori supercomputer, operated by the National Energy Research Scientific Computing Center (NERSC). KSTAR is located 
in the Daejeon in the Republic of Korea and NERSC is located in Berkeley, California. Both sites are about 9,000
kilometers apart and are connected through a network link with 100 GB/sec bandwidth, carried by Kreonet \footnote{\url{https://www.kreonet.net/eng}}
and the Energy Sciences Network (ESnet) \footnote{\url{https://www.es.net/}}. Besides high-performance computing resources,
NERSC operates container orchestration services that allow to run web-based services to support scientific projects.
Furthermore accommodates Cori's job scheduler real time quality of service for processing needs with external
real-time constraints. These amenities available at NERSC make it a useful target to deploy near real-time analysis
capabilities for remote experiments. With \DDELTA we aim to develop a framework that makes these kind of approaches easy to
implement and available for broader applications, more diagnostics.

The remainder of this article is structured as follows: In \secref{framework} we describe the design approaches and 
choices made for \DDELTA. \Secref{applications} describes in detail two example applications that \DDELTA allows to
routinely perform during tokamak operations. \Secref{benchmarks} presents performance benchmarks of \DDELTA and
\secref{visualization} describes details of the implemented web-based visualization system. Finally, \secref{conclusions}
gives conclusions and pointers for future work.

\section{The adaptable near real-time analysis framework}\label{sec:framework}
\DDELTA aims to connect fusion experiments, HPC resources, and scientists by facilitating near real-time streaming analysis
of big fusion data and provide immediate visualization of the results. This description can be translated into the following
requirements on the system. First, to move big and fast data from experiment to HPC, a robust network link is required.
Second, a streaming protocol, optimized for the needs of scientific data should be used. Third, it should be easy 
to perform user-defined data analysis kernels on the incoming data stream. Fourth, once a data packet has been
analyzed it should immediately be made available to data visualization or other downstream consumers. And finally,
access to data visualization should be readily accessible. The first requirement is a hardware constraint. The work
discussed in this contribution focuses on measurements of KSTAR plasmas on Cori, located at the National Energy Research
Scientific Compute Center. These sites are connected through the Energy Sciences Network and data transfer rate
benchmarks are reported in \cite{churchill-2021}. Second, \DDELTA uses ADIOS2 \cite{godoy-2020} as  a streaming library.
This library, which is part of the United States Department of Energy Exascale Computing Project (ECP) software technology stack
for data and visualization, allows to effortlessly transport data in a peer-to-peer manner across Wide-Area-Networks. It
implements a time-step abstractions which mimics the time-sample production cycle of  measurements at fusion experiments. 
Addressing the third prong, \DDELTA is implemented in python and implements a streaming architecture, where data kernels
are applied to individual data packets. This allows to readily include user-written kernels in any workflow. After applying
a data analysis kernel on a data packet, the results are immediately stored. This satisfies the fourth requirement. And
addressing the final requirement, \DDELTA features a web-based visualization dashboard, which can be accessed from anywhere.

\begin{figure}[htbp]
  \centering
  \includegraphics[width=\textwidth]{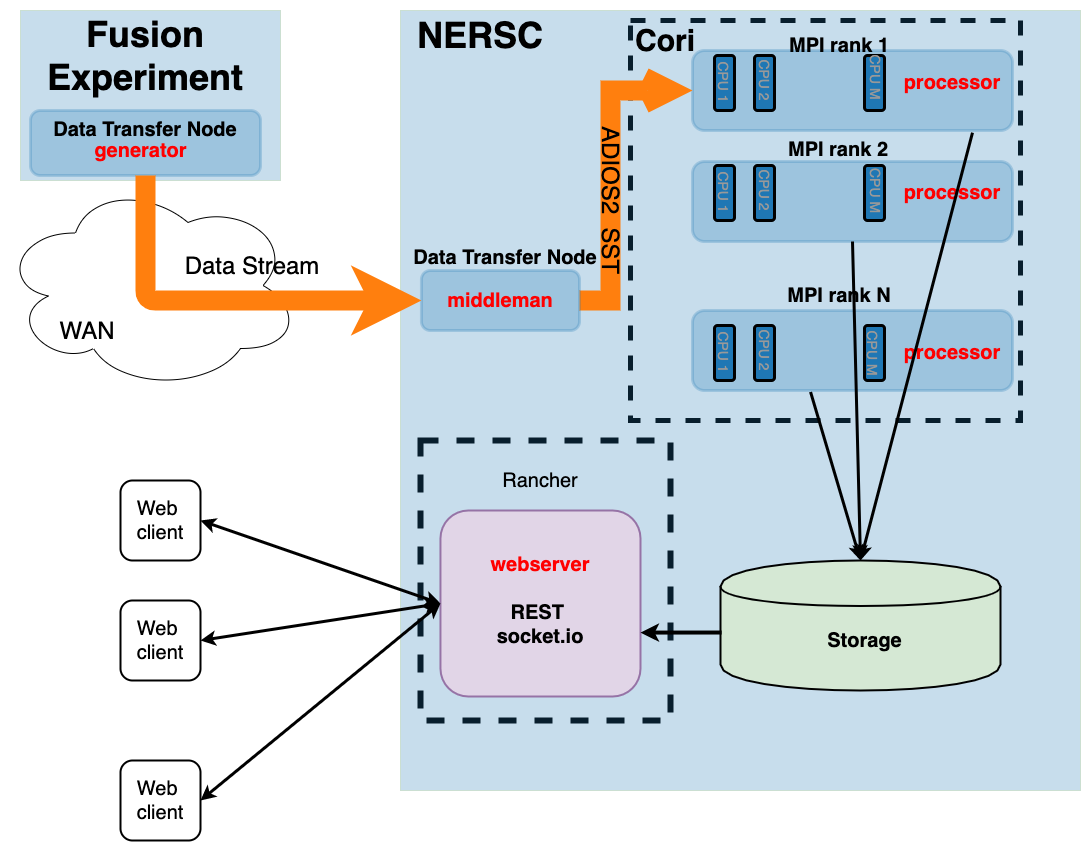}
  \caption{The architecture of the \DDELTA framework}
  \label{fig:delta_arch}
\end{figure}

To connect fusion experiments with HPC resources, \DDELTA is implemented as a distributed system. \Figref{delta_arch}
presents an overview of the systems that \DDELTA couples. At the site of the fusion experiment, a \emph{generator}
runs on a Data Transfer Node (DTN). A DTN is a server that is optimized for WAN data transfers. After an experiment
has finished, the \emph{generator} loads the measurement data, stages it for streaming, and transmits it. For this
section of the route we are using the \emph{DataMan} streaming protocol implemented by ADIOS2. At a DTN of the 
target HPC center, a \emph{middleman} relays the data stream to the target HPC resource. Employing a relay node
enables to receive the stream using the high performance network interface of the DTN. It also enables to forward
the stream to the target HPC resource using the \emph{SST} protocol which is designed for use in HPC environments.
The \emph{processor} runs as an MPI program on the target HPC resource. It receives the data stream and executes data
analysis code using distributed compute resources. The analyzed data is stored in a database immediately after it has been calculated.
This database is accessible by both, the HPC resource and externally facing services. At NERSC, the web-based dashboard runs on
Spin, a container orchestration service based on Rancher\footnote{\url{https://www.rancher.com}}. 

While the overall architecture of \DDELTA is flexible and its different components may run at systems that are networked together,
it benefits from the resources available at NERSC. For one, it makes use of distributed storage solutions that are available from both,
the used HPC resource and the externally facing container orchestration service. And it uses the container orchestration service to host
the web-based visualization service on-site. Deploying \DDELTA at HPC centers without such infrastructure is possible, but with the
additional burden of having to implement a service that moves data analysis results to the visualization service.

\subsection{Software architecture and abstractions}
\begin{figure}[htbp]
  \centering
  \includegraphics[width=\textwidth]{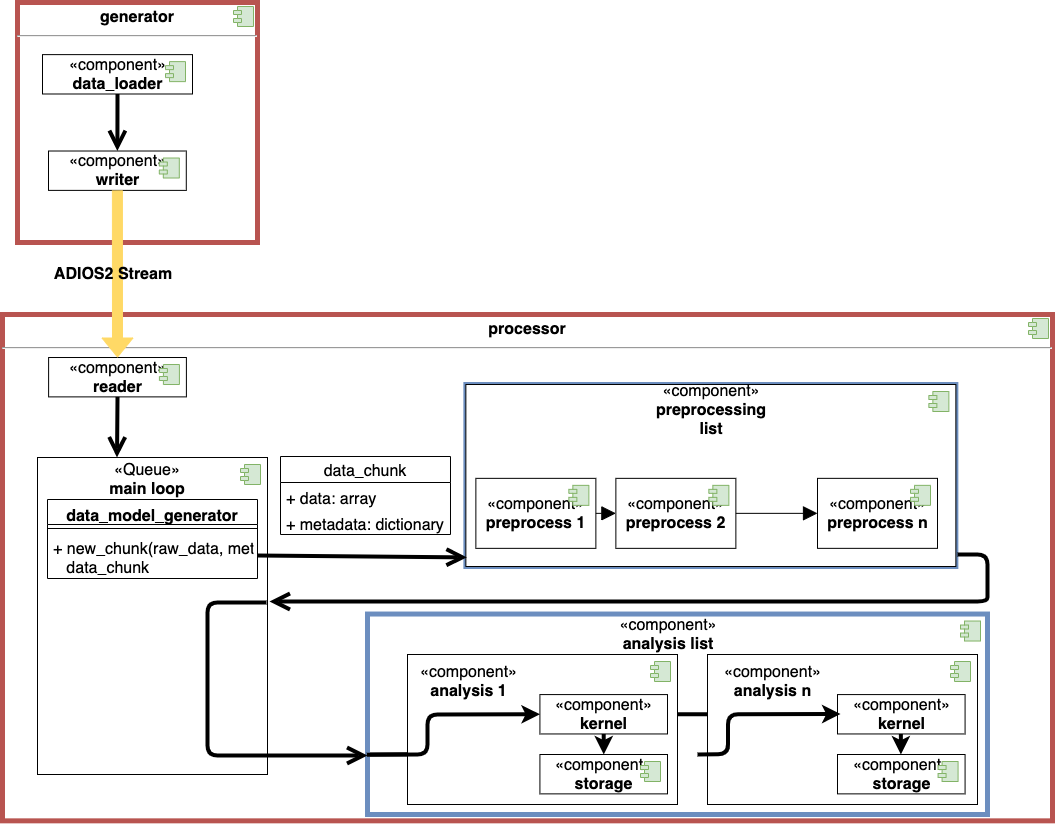}
  \caption{The software architecture of the \DDELTA framework}
  \label{fig:delta_sw_arch}
\end{figure}

To perform data analysis, \DDELTA moves data from an experiment to a HPC resource.
Figure \ref{fig:delta_sw_arch} illustrates the software architecture that facilitates this task. The goals of the
architecture are to provide abstractions to handle data from various plasma diagnostics, for data streaming,
and for data analysis workflows. These abstractions are implemented as an object-oriented design.

Data loading and staging for streaming is implemented through a \emph{data\_loader}. Methods of this class
load the measurement data together with relevant metadata from storage, such as files or local databases into memory.
Once loaded in memory, the \emph{data\_loader} exposes that data as a low-dimensional time series through an iterator
interface. The metadata of the measurements, for example sample frequency or parameters of the diagnostic, are exposed
as class attributes. At each iteration a chunk consisting of a fixed number of consecutive samples is returned, starting from the
beginning of the time series. And once the iterator is exhausted, the entire measurement time series has been iterated over.

A \emph{reader} and \emph{writer} implement data streaming. In particular, the \emph{reader}
sends data chunks, produced by the \emph{data\_loader}, as individual time-steps through an ADIOS stream. These chunks
include both, time series data as well as metadata. Once received by the \emph{reader}, the \emph{processor}
instantiates an appropriate \emph{data\_chunk} object from the time-step data, which includes both time series data
as well as metadata. This metadata may be used as parameters for the follow pre-processing and data analysis tasks.

\DDELTA implements a linear two-stage data analysis workflow that is commonly used for the analysis of fusion data.
In the first pre-processing stage, the data is subject to filters, such as smoothing or bandpass filters. These
operations are performed when measurements are subject to noise or the phenomena of interest are restricted to
certain frequency bands. A series of filters, which mutate the data in a \emph{data\_chunk} are succesively applied.
Following the pre-processing, the data analysis stage applies data analysis kernels to the chunks. These transformations
act as sinks, by storing the transformed data and returning nothing. Without downstream dependencies, the data analysis
kernels are performed in parallel. 

This linear two-stage workflow model is static and stateless, i.e. each \emph{data\_chunk} will be transformed in exactly the
same way. \DDELTA implements this workflow in a data-parallel mode by implementing a master-worker pattern in the 
\emph{generator}, centered around the incoming data stream. The main thread receives the data stream and inserts each 
chunk into a first-in-first-out (FIFO) queue. Simultaneously, a group of worker threads pop chunks from the queue.
A given thread then submits its chunk into the pre-processing queue, waits for it to return, and then to data analysis
kernels. The pre-processing and data analysis is performed on the distributed compute resources, using the mpi4py library \cite{dalcin-2021}.
In particular, \DDELTA implements the execution of the pre-processing workflow and the analysis kernels as a callable.
This allows the the worker threads to run the workflow tasks on a pool of worker processed exposed by mpi4py through a 
PEP-3148 \footnote{\url{https://github.com/python/peps/blob/master/pep-3148.txt}} compatible interface.

This architecture of \DDELTA allows to perform the entire data analysis chain, starting by reading measurements from
file and ending when the data analysis results are stored, completely in memory, across multiple computer systems.
By performing in-memory streaming, \DDELTA does not use any file-based I/O and thus avoids performance issues caused
by I/O variability \cite{xie-2017}. Another design choice is that \DDELTA uses a process pool to access distributed compute
resources. Making this design choice trades simplicity for customizability as all available MPI ranks are divided evenly
among the processes. There is no way to designate additional compute resources to more demanding analysis tasks. Finally,
the run-time configuration for \DDELTA is defined in a central configuration file. This configuration file defines what data to load,
parameters for the data streaming, the pre-processing and analysis pipelines as well as the storage backend to be sued. All components
that make use of this configuration file are denoted with a green square in \Figref{delta_sw_arch}. Given a configuration
file and the exact version of \DDELTA that was used, any data analysis can be reproduced as long as the original 
measurement data are available.

A web-server that is running on NERSCs Spin service \footnote{\url{https://www.nersc.gov/systems/spin/}} serves dashboards with plots.
While Spin is an external facing service with no direct access to HPC resources, it has access to the database into which the
data analysis kernels write. With knowledge of the database scheme, the web server can fetch the data requested by clients
and render a plot on a web site which is then delivered to the client. As the data analysis workflow, the web-based visualization
allows for multiple servers to run at the same time. Spin can seamlessly scale this service by adding web server instances
on demand. And the limiting factor for the visualization are the bandwidth from Spin to the database that stores the
analysis results.

\section{Applications} \label{sec:applications}
Operating \DDELTA along side plasma experiments allows to automatically perform large-scale data analysis and 
present informative data visualizations. This section describes two example applications for \DDELTA, one that make use of the available
compute power of the Cori supercomputer and one that uses machine learning to augment data visualizations. Both of
these applications use data produced by the KSTAR ECEi system \cite{yun-2014}.

\subsection{Automated turbulence analysis}
The largest data producer at the KSTAR tokamak is the Electron Cyclotron Emission imaging (ECEi) system. 
Sampling ECE at three different toroidal views of 24 by 8 pixels on a microsecond time scale, this system produces data at a rate of
about 1 GB/s. ECE data can be used to infer perturbations of the electron temperature \Te relative to a time-average, 
$\delta \Te / \langle \Te \rangle$. The KSTAR ECEi diagnostic maps the 24 by 8 array onto an approximately 10 by 40 cm
large view of plasma column in the radial-poloidal plane. It is thus regularly used to visualize magneto-hydrodynamic modes,
such as magnetic islands or ELM precursors \cite{park-2019}. And spectral quantities, such as the cross-power, coherence, cross-phase or
cross-correlations from channel-pair combinations of the ECEi samples, are regularly used to investigate flow phenomena
such as the interaction between magnetic islands and the background turbulence, or studying avalanche-like electron heat
transport in the core plasma \cite{choi-2017, choi-2019}. 

Using \DDELTA, a spectral analysis of core \Te turbulence can be performed automatically after each plasma shot. For this
we define the spectral cross-power $P$, the coherence $G$, the cross-phase $A$ and the cross-correlation $R$ as 
\begin{subequations}\label{eq:1}
\begin{align}
    P_{XY}(\omega) & = \mathbb{E} \left[ X(\omega) Y^{\dagger}(\omega) \right] \\
    G_{XY}(\omega) & = | S_{XY} | / \sqrt{ S_{XX} S_{YY}) } \\
    A_{XY}(\omega) & = \arctan\left( \mathrm{Im}\left(S_{XY} / \mathrm{Re}\left( S_{XY}\right) \right) \right) \\
    R_{XY}(\omega) & = \mathrm{IFFT} \left( S_{XY} \right),
\end{align}
\end{subequations}
where $\omega$ denotes angular frequency, $\mathbb{E}$ denotes the expected value operator, $^\dagger$ denotes complex conjugation,
and $X$ and $Y$ denote Fourier-transformed time series. These quantities are regularly used in analysis of turbulent heat flows in the core
of fusion plasmas, and provide information on local dispersion relation and have also been used to identify avalanche-like $\Te$ transport events
\cite{choi-2017, choi-2019}.

The ECEi diagnostic samples cyclotron radiation using an array of $24$ by $8$ pixels so that there are $\binom{192}{2} = 18336$ possible
channel pair combinations for which to calculate \Eqnref{1}. Typical pulse lengths at KSTAR are 5 seconds, in which the ECEi samples $5,000,000$ samples.
This time series is partitioned into $500$ time-chunks, of $10,000$ samples by the \emph{generator}. In a streaming setting, any calculation is
performed individually on each time-chunk. After normalizing each chunk, the chunks are streamed to the \emph{processor}.
There, each chunk is Fourier transformed by a pre-processing routine and subsequently analysis kernels
that implement \Eqsref{1} are applied to the pre-processed chunk. 

This computationally demanding workflow is an exemplary workflow which can be automatically performed using \DDELTA.
It automatically provides researchers with detailed analysis of core plasma turbulence. Mining such data automatically
for each plasma pulse would allow compile a large database which could be of use for pattern recognition algorithms.
Alternative use-cases for \DDELTA would be integrated modeling simulations based on time-slice measurements or
tomographic inversion of bolometry data. For the remainder of the article we use the ECEi turbulence workflow as a
benchmark to characterize the performance of the \DDELTA framework and identify bottlenecks that arise as a consequence
of the implemented architecture.

\subsection{Automated image analysis}
Automatic image analysis and visualization is another task that is readily implemented using the \DDELTA framework.
Again, we consider ECEi data, whose spatial sampling of \Te fluctuation has been crucial to identify details
of MHD instability physics in KSTAR plasmas \cite{park-2019}. For example, ECEi data has been used in combination
with a simplified tearing mode model to estimate stability parameters of magnetic islands \cite{choi-2014}. As an
example of how automatic image analysis and visualization may be incorporated in \DDELTA we consider the task of
identifying magnetic island structures in ECEi data.

Magnetic islands are located around rational field lines and are structures where the topology of a fusion plasma
is not a set of nested flux surfaces. Instead, a magnetic island appears as flux surface that has been split
open into an elliptic structure, with a half-axes oriented along the radial and poloidal directions. This splitting 
of the field lines is associated with magnetic reconnection and magnetic islands are associated with resistive MHD
instabilities, such as tearing modes \cite{waelbroeck-2009}. With one half-axis of the magnetic island oriented along the radial
direction, the perturbed magnetic field lines span a range of radial positions. Electrons that follow the perturbed
field lines can then quickly transport heat across this range of radial positions. Magnetic islands can thus flatten
the electron temperature profile once they are sufficiently large. The footprint of this effect in ECEi data are
so-called radial phase inversion structures, quadrupole-like coherent patches in the local \Te measurements \cite{choi-2016}.

\begin{figure}[htb]
    \centering
    \includegraphics[width=0.9\textwidth]{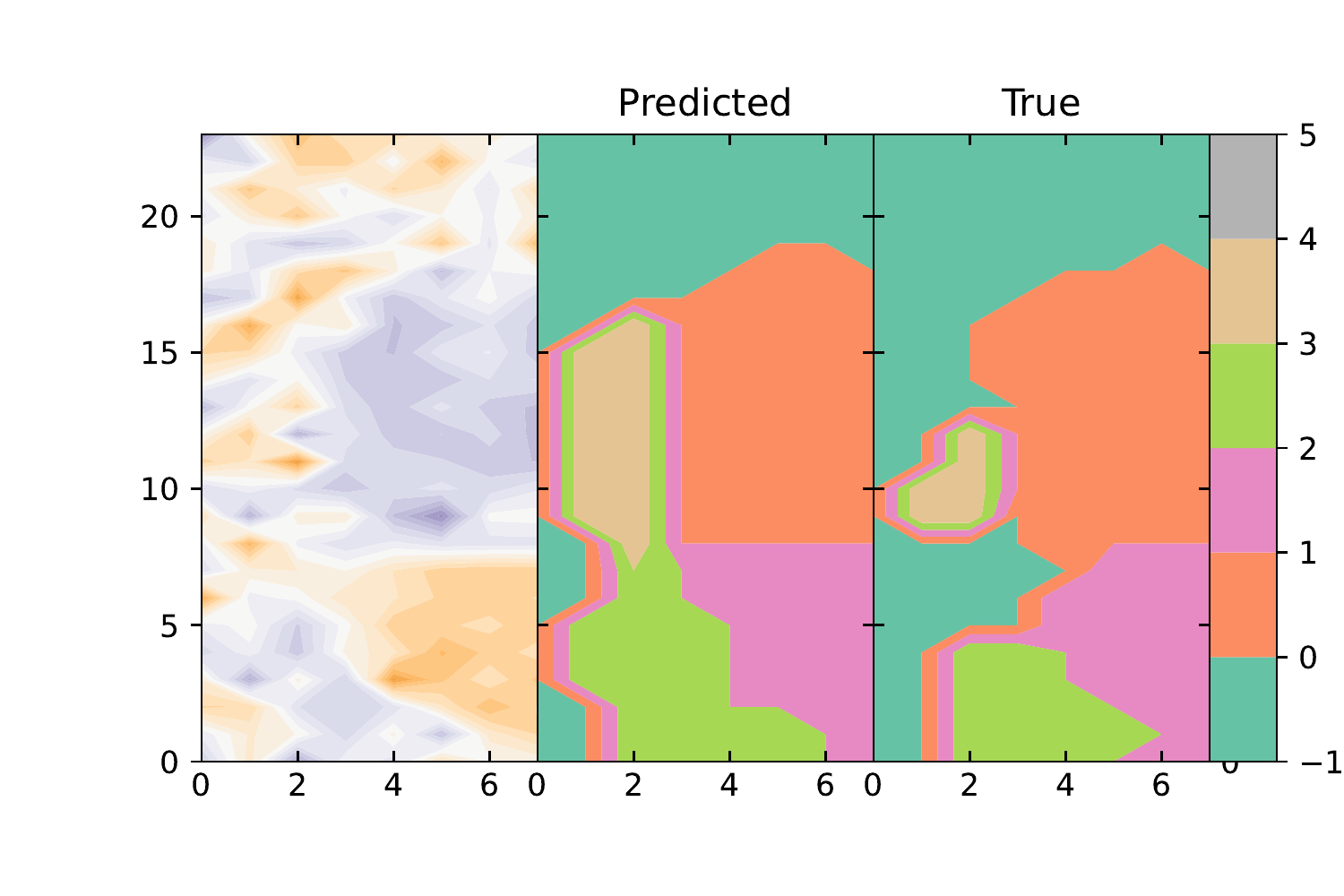}
    \caption{ECEi data showing a quadrupole-like footprint of a magnetic island (left). Segmentations of the 
    image, where 0 denotes background and 1 through 4 denote the different regions of the quadrupole structure are shown
    in the middle and the right plot.}
    \label{fig:semantic_segmentation}
\end{figure}

Image segmentation is a computer vision task that identifies structures in images. In particular, semantic segmentation
assigns each pixel a label that identifies it with a set of pre-determined structures. This task can be applied to
ECEi data to identify radial phase inversion structures associated with magnetic islands. Figure \ref{fig:semantic_segmentation}
illustrates this principle. The leftmost contour plot shows $\delta \Te / \langle \Te \rangle$ and the radial phase inversion structure
occupies the lower 80 percent of the image. There are two prominent peaks of opposite sign spanning rows 4 to 7. Left from
those peaks are structures with opposite phase, hence the name radial phase inversion structure. Segmentations of this image
using 5 labels is shown in the other two. Here labels 1 through 4 denote the individual peaks of the radial phase inversion
structure and background pixels are labelled with 0. Since the measurements are subject to noise the manually provided labels
are always subject to biases.

A well-established model for semantic segmentation tasks is the U-Net model \cite{ronneberger-2015}. It consists of a symmetric
encoder-decoder structure where the encoder is comprised of a series of convolutional and pooling layers. This is a common
architecture in convolutional neural networks. The decoder then expands the encoded information by applying the same
number of transpose convolutions until the data has approximately the same size as the input. In addition, skip connections propagate
information from the different stages of the encoder to stages of appropriate size in the decoder. Since ECEi data is given as 24 by 8 pixel
images, the original U-Net architecture needs to be dramatically simplified in order to work with this data. 
The adapted architecture first applies two $3 \times 3$ convolution to the data. The first convolution increases the image from $1$
to $8$ channels and is followed by applying a rectified linear unit (ReLU). After the second convolution, a $2 \times 2$ max pooling
operation is performed. The second and third convolutional blocks in the encoder apply the same operations and increase the
number of channel from $8$ to $16$ to $32$. In the expansive decoder part, the input is first subject to a $2 \times 2$
up-convolution and then concatenated with the output of the encoder block of matching size along the channel dimensions.
This data is then passed through two convolutional layers using $3 \times 3$ filters and a padding of 1 in order to preserve
image resolution. The final convolutional layer maps the previous layer onto the number of pixel classes, $5$ in our case.
This architecture has been implemented using pytorch \cite{pytorch}, the source code is available here \cite{kube-ecei-ml-utils}.

\begin{figure}[htb]
    \centering
    \includegraphics[width=0.5\textwidth]{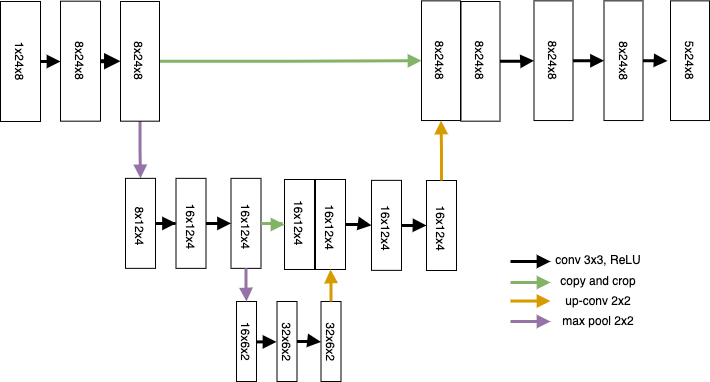}
    \caption{U-Net architecture}
    \label{fig:unet}
\end{figure}

We compiled a training set by manually segmenting $350$ images with visible radial phase inversion structures. To train the network we
augment this set using random crops of size $(12,4)$, $(20,6)$, and $(16,6)$ pixels, rescaled to the original resolution, as well as
with a set of duplicate images where random noise has been added. This yields in total $2,100$ training samples, $80$ percent of these 
images are used for training and $20$\% are used as a validation set. We use the ADAM optimizer \cite{kingma-2017-adam} to optimize a categorical
cross entropy loss function \cite{good-1952}, which is commonly used for semantic segmentation tasks.
This loss function calculates a per-pixel loss between the correct label and the label probabilities predicted by the network.
In order to interpret the network output as a probability, a softmax function is applied pixel-wise to the final layer.
Starting with a learning rate of $10^{-3}$, we find that the per sample loss on a test set is minimized after about
$15$ epochs.

\begin{figure}[htb]
    \centering
    \includegraphics[width=0.5\textwidth]{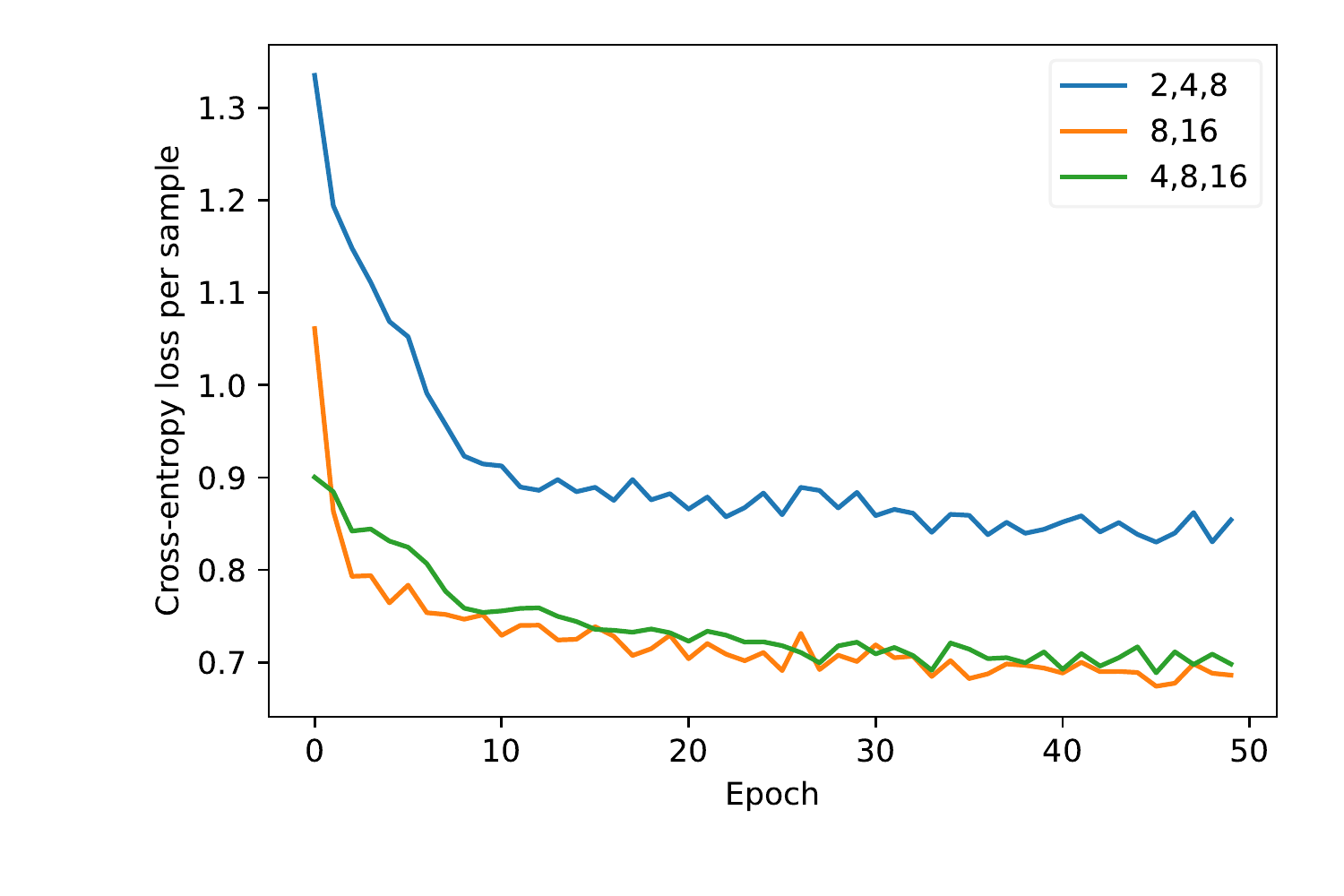}
    \caption{Validation loss of the reduced U-Net model for different architectures. The training loss curves follow the
    validation loss curves shown here tightly.}
    \label{fig:ml_loss}
\end{figure}

In order to identify optimal parameters for the architecture we varied the number U-Net layers, the number of channels in the
convolutions, the batch size used for training, as well as the optimizer. Empirically we find that the model performs well
as long as one uses at least two convolutional layers in the U-Net architecture, and at least 16 channels. More shallow
U-net architectures perform worse. This is shown in \Figref{ml_loss}, where we plot the per-sample loss calculated over
the validation set over the training of three different U-Nets. The first one has three convolutional layers, respectively with
2, 4, and 8 channels. The second one has two convolutional layers, with 8 and 16 channels, and the third U-Net has three layers,
with 4, 8, and 16 channels. Just increasingthe number of convolutional channels from 8 to 16 decreases the per-sample loss by 
about 30\%. When varying the mini-batch size, the results in \Figref{ml_loss} are obtained using a mini-batch size of $16$, but found
only negligible effect on the loss. Stochastic gradient descent with a learning rate of $10^{-3}$ performed on-par with the ADAM optimizer
when we used a learning rate scheduler that divided the learning rate by two on plateaus.

The forward pass for such simple models requires only little computational power, the models discussed above have between $5$ and $10,000$
parameters. Thus, this model van be seamlessly integrated into the server that serves visualization dashboards, as discussed
in \secref{visualization}. Such machine learning models can also be used in the pre-processing stage. By adding image
segmentation information to a chunks metadata, they could target data analysis tasks to certain parts of the image.

\section{Performance Benchmarks}
\label{sec:benchmarks}
\DDELTA is a complex distributed system that runs on high-performance computing resources to analyze data streams in near
real-time. Even though it uses traditional HPC libraries and follows a traditional single program multiple data
paradigm, their usage pattern in \DDELTA differs from traditional HPC software, such as numerical simulation codes. Particularly, \DDELTA
implements a \texttt{master-worker} architecture where the main loop distributes pre-processing and analysis tasks to a
large number of workers. Workers are instantiated on demand as separate threads or processes and execute user defined pre-processing and analysis
tasks, which could by frequency filters, short-time Fourier transformations,
coherence functions for signals, or even tomographic inversion analysis of bolometry data \cite{ingesson-1998, montisci-2021}.
All of these examples have widely different algorithmic complexity, such as $N \log N$ for the Fourier transformation,
$N$ for coherence functions, assuming that the data has been Fourier transformed, and algorithms that implement such analysis
often perform best using different memory access patterns.
\DDELTA aims to facilitate the performant execution of a broad range of algorithms in a streaming setting, where data analysis
tasks may change from one run to another. And optimizing the framework for the execution of certain tasks may lead to
performance bottlenecks for other tasks. Thus, \DDELTA distributes all available resources equally among the workers.
This is in contrast to traditional HPC software for numerical simulations, where a-priori knowledge of the algorithms, 
memory access patterns, sequence of operations, and the data can be used to optimize execution on any given hardware.

\subsection{Individual components}
As a first step in understanding the performance of \DDELTA we evaluate the performance of its individual components in
a synthetic setting. In the following we explore the performance of the network connection between KSTAR and NERSC,
the filesystem on NERSC as well as implementation of Eqs.(\ref{eq:1}) for graphical processing units (GPUs). 

\begin{figure}[htb]
    \centering
    \includegraphics[width=\textwidth]{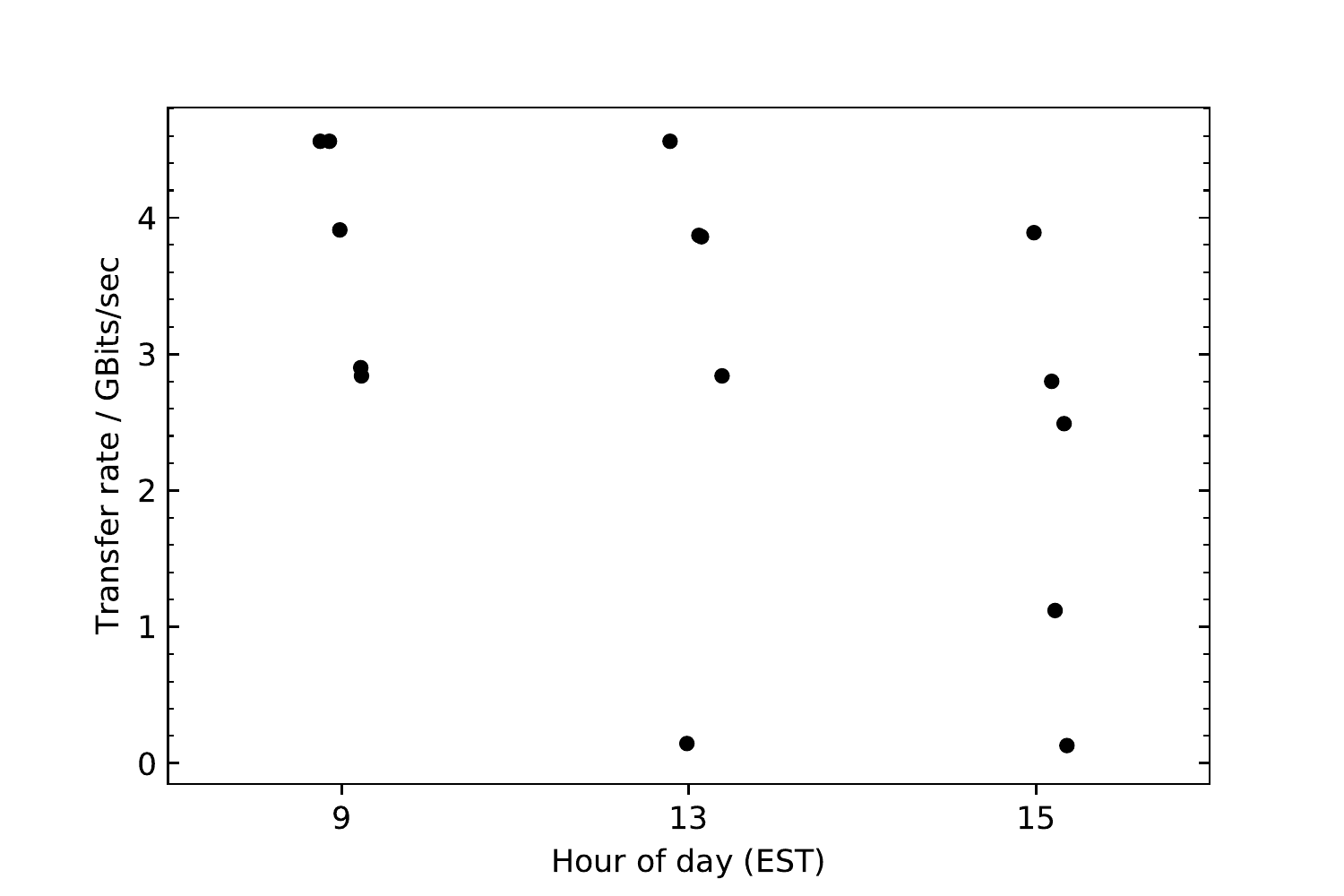}
    \caption{Transfer rates between the DTNs at KSTAR and NERSC, measured using iperf.}
    \label{fig:iperf_transferrates}
\end{figure}

Network traffic originating from the KSTAR DTN is routed through Kreonet and ESnet to NERSC, on a shared 100 GB/sec link. As a shared link,
the achievable bandwidth is subject to fluctuations between streaming events. In order to establish a baseline of the expected
bandwidth and to characterize link variability we measured the achieved bandwidth in multiple runs of the iperf benchmarking tool
\footnote{\url{https://iperf.fr/}} over one week in August 2021. As shown in \Figref{iperf_transferrates}, the link consistently
sustains data transfers at rates over $3$ Gbits per seconds. At several occasions we observe that the link peformance was
degraded, which may be due to traffic congestion from other applications. We do no observe a dependence of the sustained
data transfer rate on the time of day.

We continue by evaluating the performance of computational kernels implementing Eqs.(\ref{eq:1}). Previous work showed that
kernels which implement Eqs(\ref{eq:1}) for central processing units (CPUs) can utilize multithreading and scale linearly
with the numbers of threads. In this work we utilize a kernel that fuses calculations of Eq.(1a), (1b), and (1c) for a
a given channel pari $(X,Y$) and runs on GPUs. We implement this kernel using the CUDA interface for numba \cite{numba}.
The kernel launch can be configured using up to 1024 hardware threads. Running on Cori we measure the kernels runtime as
approximately $0.8\,\mathrm{s}$ when launching with $(32,32)$ threads. For $(16, 16)$ and $(16, 32)$ threads we observe changes in
runtime less than $3\%$ and therefore set $(32,32)$ as the default number of threads used in the remainder of this article.
Equation (1d) is implemented by a separate kernel and also runs on the GPU. But as it only consists of a inverse Fourier
Transformation it has no free parameters and we report on its performance only in the context of a complete data analyis
workflow in the following subsection.


\subsection{Cori scaling and performance}
To evaluate the performance of \DDELTA, we perform the benchmark ECEI workflow described in \Secref{applications} on
the newly added GPU partition on Cori. This partition consists of 18 nodes, each with two 20-core Intel Xeon Gold 6148 CPUs,
384 DDR4 memory and 8 NVIDIA Tesla V100 GPUs, each with 16 GB HBM2 memory. These nodes are networked through 4 dual-port
Mellanox MT27800 EDR InfiniBand HCAs. The benchmark workflow consists of calculating a short-time Fourier Transformation
of each data chunk, followed by evaluating \Eqsref{1}.  We configure \DDELTA to execute as an MPI program where the root
process executes the main loop and serves as a master process for all child processes. Spawned from the master process,
a number of worker threads submit individual time chunks into a pre-processing queue and finally into a analysis queue.
The pre-processing routines are executed as individual threads by the root process. While this limits the number of time chunks that can be pre-processed
simultaneously it avoids MPI communication overhead as threads can access the time-chunks through shared memory. This is advantageous
given the small size of the time-chunks. The analysis tasks are distributed among the MPI ranks. Each MPI rank executes a fused calculation of
Eqs.(1a) - (1c) into a single kernel as well as a kernel that implements Eq.(1d) on a GPU. Table \ref{tab:param_scan} lists
the number of worker and pre-processing threads as well as the number of MPI ranks that were used for the analysis.

\begin{table}[htbp]
    \centering
    \begin{tabular}{c|c}
        Resource                & Number of parallel execution units \\ \hline
        Queue worker threads.   & 1, 2, 4 \\
        Pre-processing threads  & 1, 2, 4, 8 \\
        Analysis MPI ranks      &  1, 2, 4, 8, 16, 32
    \end{tabular}
    \caption{Number of parallel execution units that were made available to queue dispatching, pre-processing, and data analysis in the
    parameter scan.}
    \label{tab:param_scan}
\end{table}

\begin{figure}[htbp]

\subfloat[Allocation A]
{
    \includegraphics[width=0.45\textwidth]{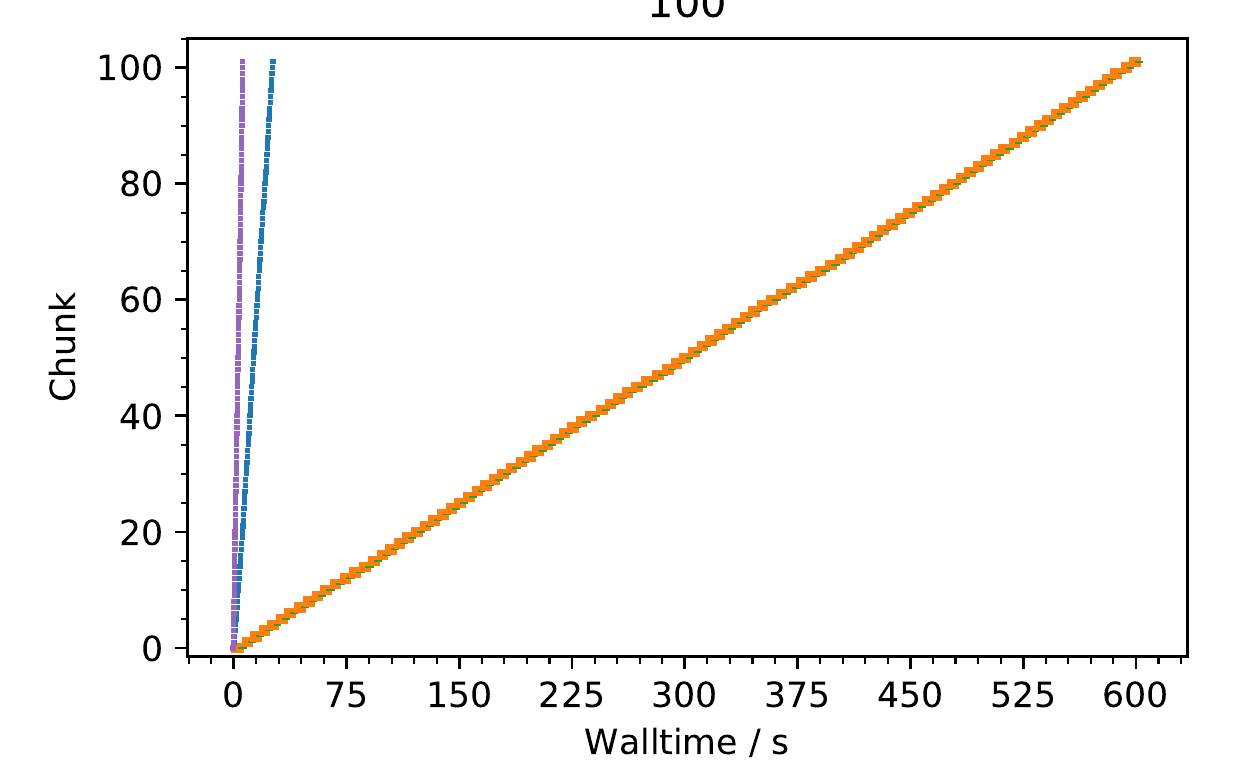}
    \label{fig:exec_timeline_1}
}
\subfloat[Allocation B]
{
    \includegraphics[width=0.45\textwidth]{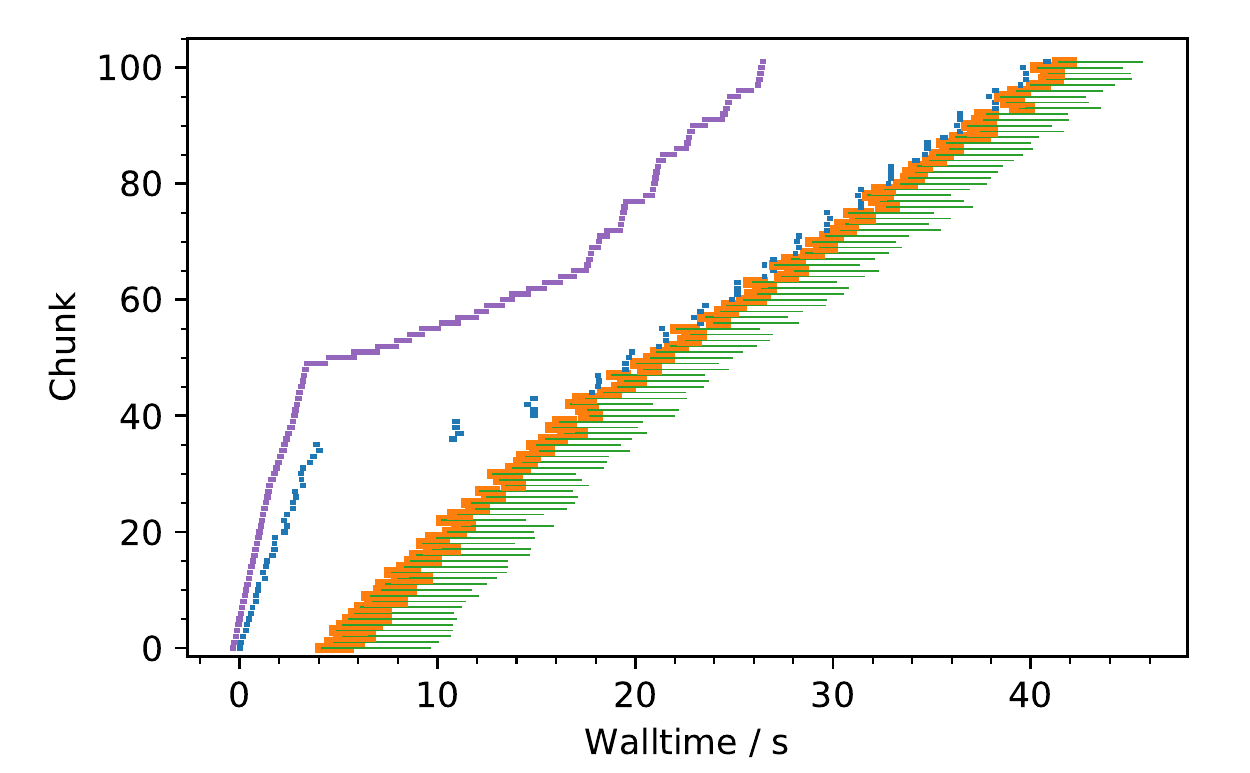}
    \label{fig:exec_timeline_2}
}
\label{fig:exec_timelines}
\caption{Execution timeline for the benchmark turbulence workflow using two different resource allocations. Blue horizontal lines
extend over the time slice where a given chunk was pre-processed. The index of the chunk corresponds to the y-coordinate of each line.
Orange and green lines respectively extend over the time intervals where the GAP and cross-correlation kernels were applied to a time
chunk. The purple lines extend over the time interval starting where it was loaded from disk and ending when it was passed to the
pre-processing list.}
\end{figure}

The \DDELTA processor allows to specify the number of queue worker threads that pop time-chunks from the main-loop's queue, the number of
threads that execute pre-processing tasks, as well as the number of MPI ranks to which analysis tasks are assigned. To understand 
how well additional resources are used by \DDELTA we define two resource allocation scenarios, one sparse and one plentiful.
Allocation A includes a single queue worker task to pop time chunks from the main loop's queue and submit them to the pre-processing
and data analysis pipelines, a single thread to execute the pre-processing pipeline and a single MPI rank with a dedicated GPU, to execute 
the data analysis kernels. Allocation B includes 4 queue worker threads, 8 threads for pre-processing and 32 MPI ranks for data analysis.
As shown in Figures (\ref{fig:exec_timeline_1}) and \ref{fig:exec_timeline_2}, providing an increased number of computing resources reduces
the walltime required to perform the benchmark workflow. Running within allocation A, \DDELTA performs the benchmark workflow
in about 300 seconds. While all 100 time chunks had been pre-processed after about $20$ seconds, their subsequent analysis took about
300 seconds to execute. Figure \ref{fig:exec_timeline_1} also shows that all time chunks are processed in-order, that is, in the
order in which the respective pre-processing and analysis tasks are submitted. Running within allocation B reduces
the required walltime to about $45$ seconds, as shown in \Figref{exec_timeline_2}. Time chunks are immediately pre-processed and data analysis
does not start until about 5 seconds into the run. We observe that this causes a delay for pre-processing. And similar as with resource allocation
A do all compute tasks, pre-processing and data analysis, begin execution in regularly spaced time intervals. In other words, the selected
software architecture facilitates robust execution of incoming data time chunks on the allocated distributed compute resources. 
We do not observe any behaviour of the compute resource that blocks the execution of the streaming data analysis.

\begin{figure}[htbp]

\subfloat[Allocation A]
{
    \includegraphics[width=0.45\textwidth]{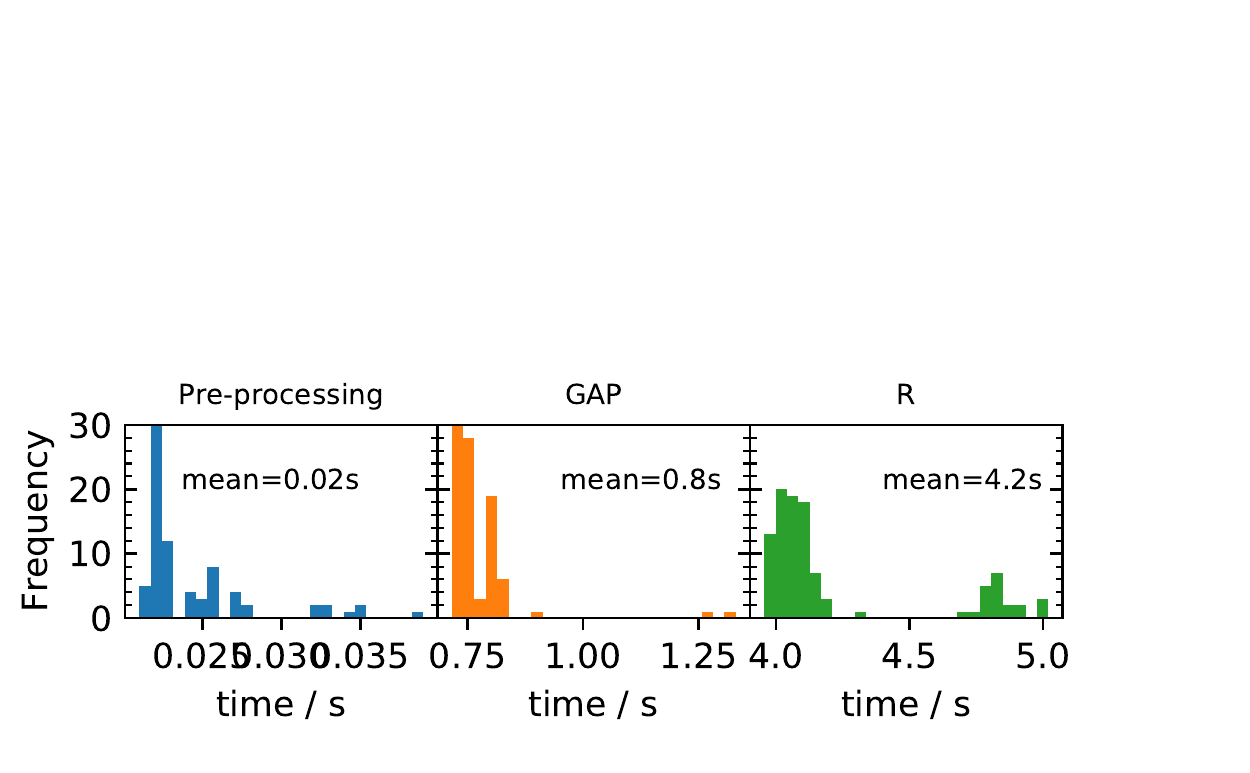}
    \label{fig:exec_histo_1}
}
\subfloat[Allocation B]
{
    \includegraphics[width=0.45\textwidth]{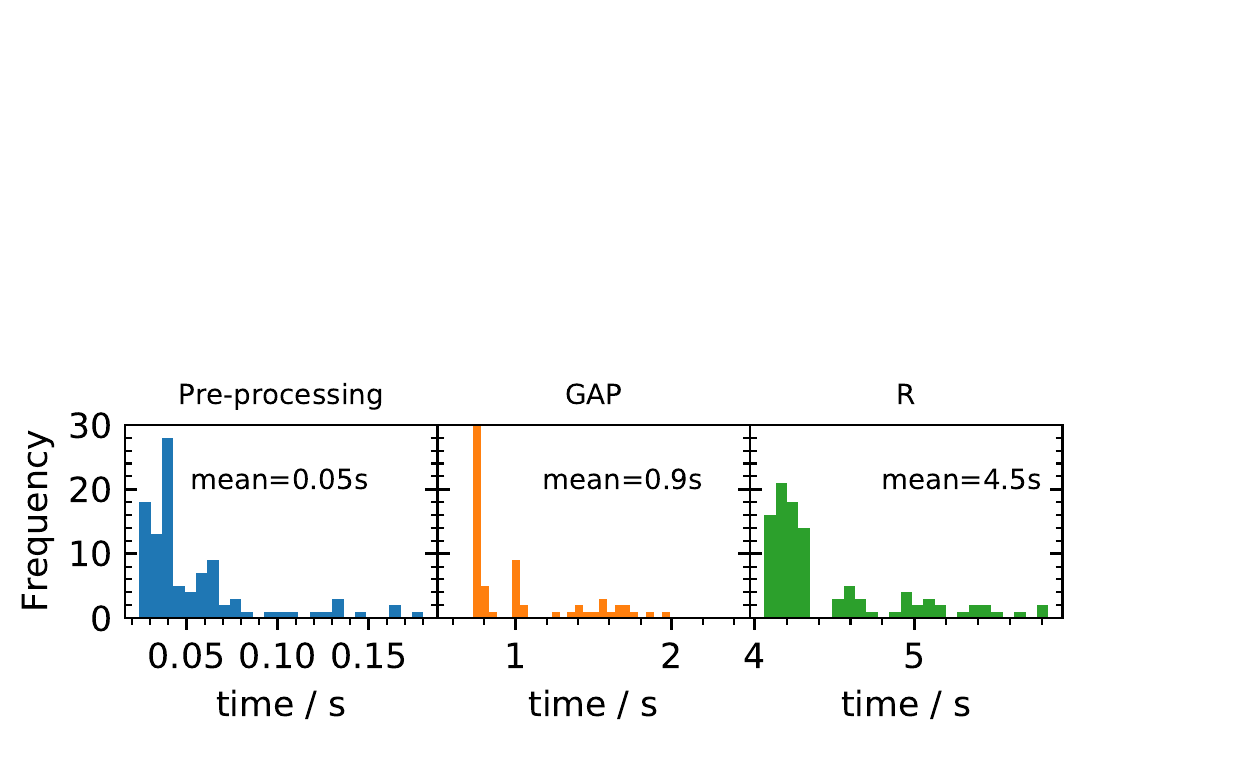}
    \label{fig:exec_histo_2}
}
\caption{Histograms for measured run-time of pre-processing and data analysis kernels in the benchmark workflow.}
\label{fig:exec_histograms}
\end{figure}

We continue by investigating the performance of the individual compute tasks, pre-processing and data analysis, when launched using
resource allocations A and B. Figure \ref{fig:exec_histograms} shows frequency counts of their measured execution time, which form
a uni-modal histogram for both, allocations A and B. When executed under allocation B, we observe a larger spread in the execution time.
This may be caused by circumstantial factors that arise when a compute resource is subject to higher load, such as increased network and I/O loads.
But on average the execution time for individual compute tasks does not vary significantly.

\begin{figure}[htbp]
    \centering
    \includegraphics[width=\textwidth]{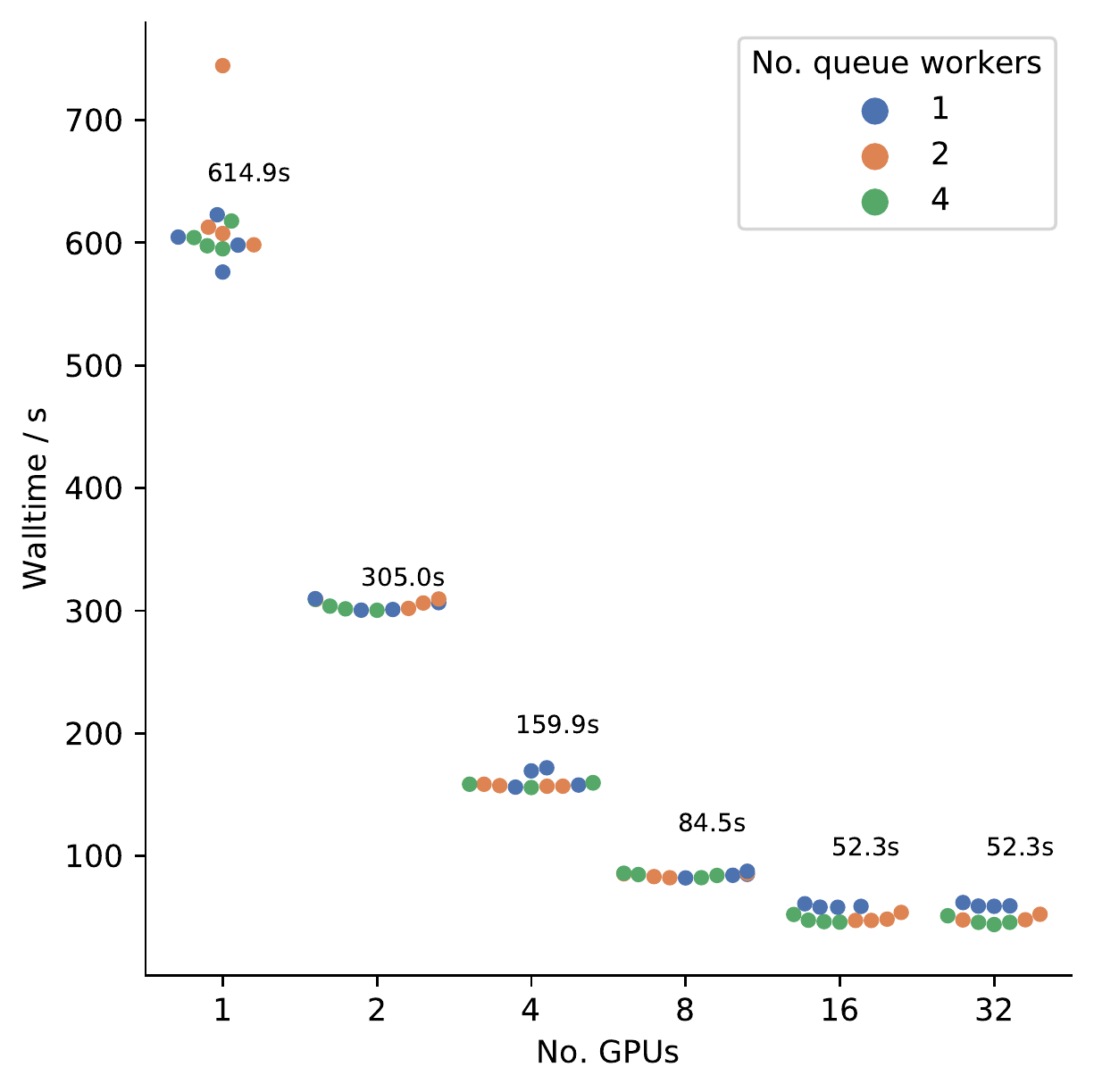}
    \caption{Wall-time to execute the benchmark workflow using the resource allocations listed in \Tabref{param_scan}.}
    \label{fig:walltime_paramscan}
\end{figure}

Finally, Figure \ref{fig:walltime_paramscan} shows the wall time it takes to execute the benchmark workflow for resource
allocations given by all possible combinations of the parameters listed in \Tabref{param_scan}.
The data point cloud centered around the number of GPUs available denotes runs with
varying number of queue workers, also encoded in color, and number of threads available to pre-processing and the average of all
walltimes is given next to each point cloud. We find that varying either affects the walltime only marginally. Running on a single GPU,
\DDELTA executes the benchmark workflow in about $600$ seconds. Doubling the allocated GPUs decreases the average walltime by half down
to about $52$ seconds on allocations up to $16$ GPUs.  Adding further compute resources to the allocation does not decrease the walltime any further.
This limit of about $50$ seconds is due to the time it takes to pre-process the data. As configured for the benchmark runs, \DDELTA performs
pre-processing using multiple worker threads on the root node. Limited by the hardware threads of the CPU, additional
hardware resources available to execute data analysis kernels idle while waiting on pre-processed data.

System dependent bottlenecks on Cori include the performance of the mpi4py library given the used MPI implementation,
as well as filesystem I/O issues that arise in large-scale python application \cite{stephey-2021}. In order to implement
a master-worker architecture, \DDELTA relies on spawning new MPI processes at runtime. The performance of these
operations may differ between implementations of the MPI standard. Second, it is observed that the performance
of python MPI applications degrade on Cori, caused by simultaneous file system requests \cite{stephey-2021}. This effect
can be mitigated by launching the requested resource configuration using a container image as the root partition.
We investigated these effects by benchmarking the turbulence benchmark workflow with \DDELTA running as a containerized
application. For this we used NERSCs \emph{shifter} service. While \DDELTA inside the container image uses MPIch,
the uncontainerized version uses OpenMPI. Scanning over all possible combinations of the parmaters listed in 
\Tabref{param_scan}, \DDELTA running as a containerized application using \emph{shifter} only shows minor differences
to the walltimes reported in \Figref{walltime_paramscan}. This implies that neither implementation details of
MPI runtime process spawning nor blocking file system requests pose a significant bottleneck for \DDELTA running on Cori.

\subsection{Performance in a streaming setting}
Finally, we benchmark the performance of the \DDELTA \emph{processor} in a streaming setting. Here, data is 
streamed from the KSTAR DTN to the NERSC DTN. There it is forwarded to the \emph{processor}, which executes
the turbulence benchmark workflow using resource allocations A and B as before. In our experiments we observe
transfer speeds of about $500$ MByte/sec from KSTAR to NERSC, which is comparable to the maximal speeds
recorded in the network link benchmark \Figref{iperf_transferrates}. Thus the ADIOS2 library uses the 
available network bandwidth optimally in this setting. At this rate it take about $5.6$ seconds to transfer
the approximately $7$ GByte of data from KSTAR to NERSC.

\begin{figure}[ht]

\subfloat[Allocation A]
{
    \includegraphics[width=0.45\textwidth]{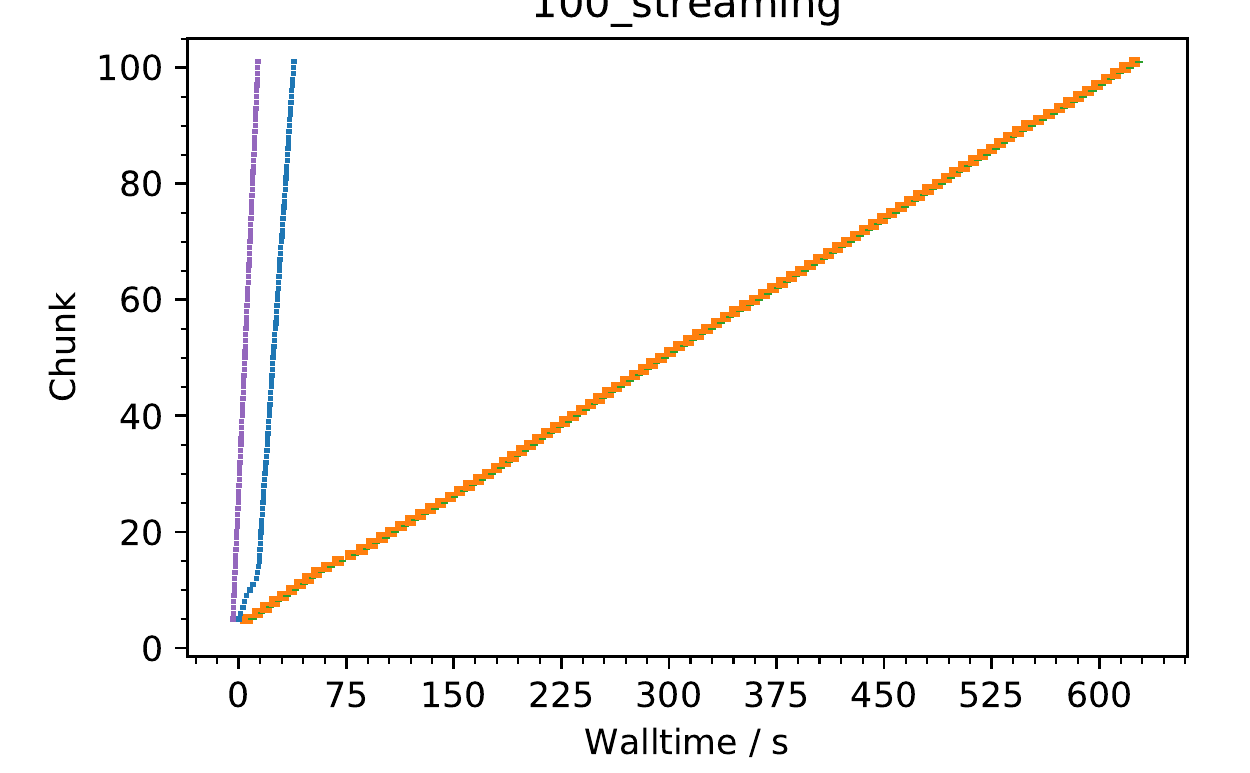}
    \label{fig:exec_timeline_1_streaming}
}
\subfloat[Allocation B]
{
    \includegraphics[width=0.45\textwidth]{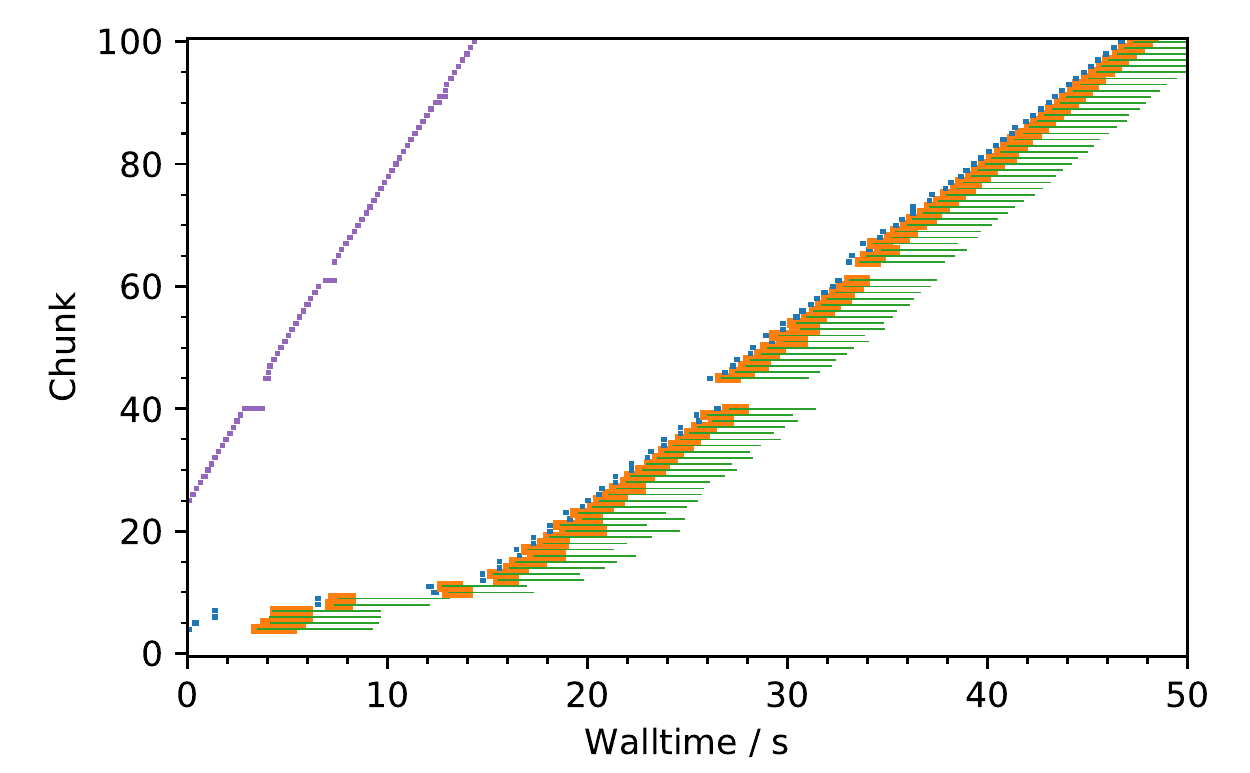}
    \label{fig:exec_timeline_2_streaming}
} 
\newline
\subfloat[Allocation A]
{
    \includegraphics[width=0.45\textwidth]{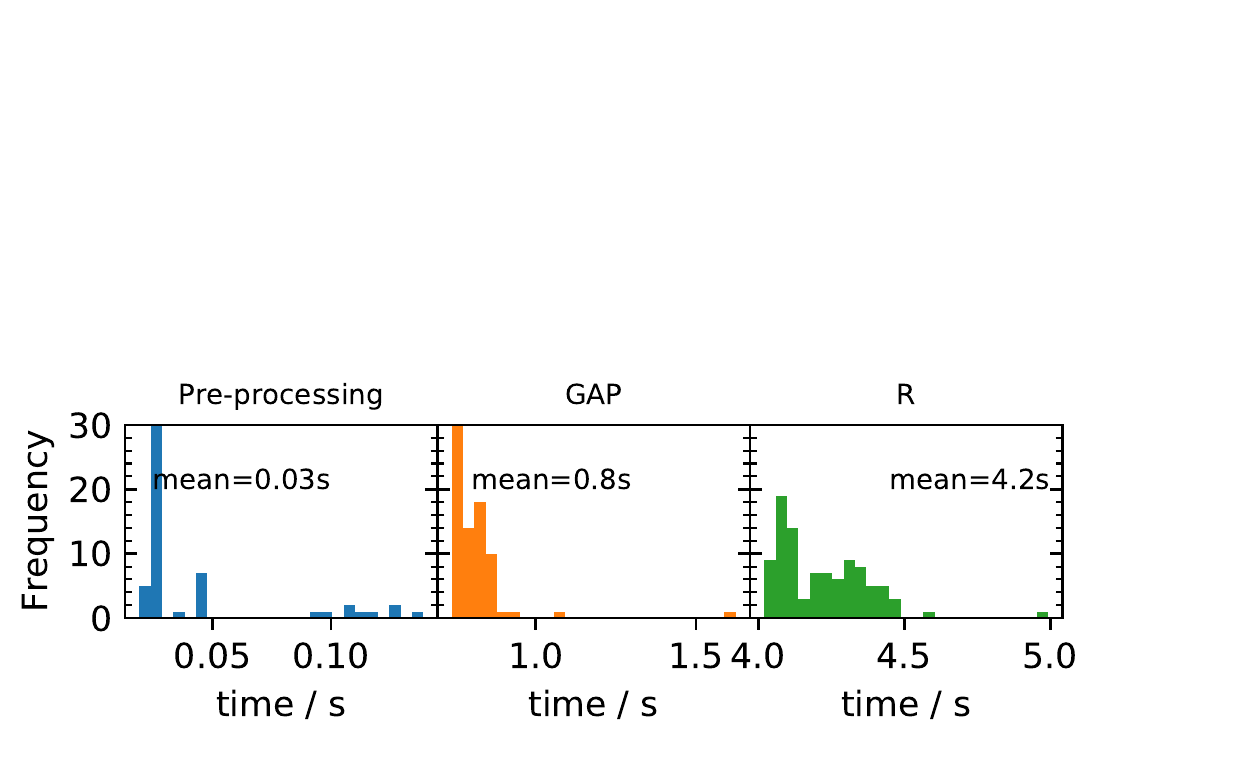}
    \label{fig:exec_histo_1_streaming}
}
\subfloat[Allocation B]
{
    \includegraphics[width=0.45\textwidth]{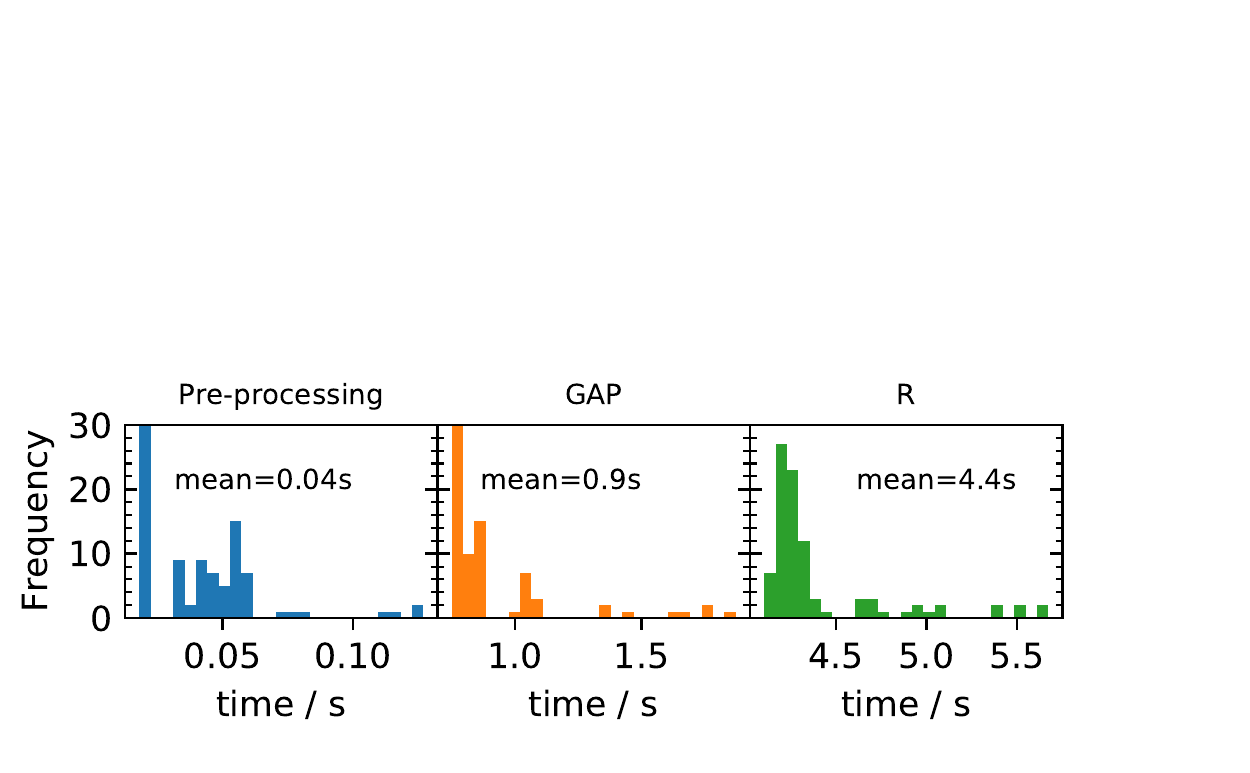}
    \label{fig:exec_histo_2_streaming}
}
\caption{The upper row shows the execution timeline for the turbulence workflow in a streaming setting 
and the lower row shows histograms of the measured run-time of pre-processing and data analysis kernels.
The left and right column show \DDELTA using resource allocation A and B respectively.}
\label{fig:performance_streaming}
\end{figure}

Figure \ref{fig:performance_streaming} shows the execution timeline of the different pre-processing and data analysis kernels
as well as histograms of measured wall-times when running \DDELTA in the streaming setting. Using resource allocation A,
the \emph{processor} finishes the workload in about $600$ seconds, similar to the time measured in a non-streaming setting.
While the total walltime to perform the benchmark workflow using resource allocation B is about 50 seconds, also the same
as measured in a non-streaming setting, the time when the individual compute tasks are performed varies. First, the
pre-processing of a given chunk is now performed immediately before any data analysis kernel is executed. In a non-streaming
setting we observed this behaviour only after the first 40 chunks had been processed. We also observe a small delay where no
data seems to be analyzed occurring after the first 10 chunks are processed. After that, pre-processing and data analysis kernels
are executed at a uniform rate until the end of the data stream. Running \DDELTA in a streaming setting does not affect the
walltime required to perform individual compute tasks. As shown in the histograms \ref{fig:exec_histo_1_streaming}
and \ref{fig:exec_histo_2_streaming}, the average time to execute pre-processing and data analysis kernels, as well as
the observed distribution is similar to those observed in the non-streaming setting.

Overall, we observe that \DDELTA executes the turbulence benchmark workflow with the same performance, regardless whether
the data is read from disk or streamed from a fusion experiment. Pre-processing of the individual data chunks bounds the
total execution time, this task is bound by on-node compute capability. The benchmark workflow is embarrassingly parallel
and \DDELTA scales strongly when increasing the number of available GPUs from 1 to 16. Adding more GPUs to the
allocation exposes the pre-processing limit. Run-time of the individual compute units, pre-processing and data analysis
kernels, is found to be independent of the available compute resources. As an archetypical data analysis task, the results
from this benchmark may translate to other data analysis tasks performed with \DDELTA.

\section{Visualization}
\label{sec:visualization}
A primary motivation for developing \DDELTA is to provide fusion scientists with data analysis results immediately after
a discharge. \DDELTA implements web-based visualizations, which allows to access them from across the globe. To implement this service, 
\DDELTA makes use of multiple services provided at NERSC. First, the \emph{processor} stores data analysis results in a locally running
MongoDB \footnote{\url{https://www.mongodb.com}} instance which is directly accessible from the compute nodes. And second, the web-server that
serves the visualization dashboards runs on a container orchestration service. Running within this service, the web-server has direct
access to MongoDB. In addition, container orchestration services such as Rancher allow to seamlessly meet workload demands. New instances
of the web server can be instantiated if too many web clients connect at once for example. In this setup, the dashboards can visualize data as
it is being analyzed by the \emph{generator}. Access restriction mechanisms that can be implemented for this setup include common control
mechanisms, such as user authentication handled by the web server, transport layer security such as the use of hypertext transfer protocol secure
(HTTPS), or restrictions based on IP addresses.

Figure \ref{fig:delta_arch} illustrates how the components of \DDELTA are used to implement the visualization service.
A web server runs on NERSCs Rancher service and forwards data requests from web clients to the MongoDB instance.
User clients can request individual data, or pull, data. This method is implemented
using a REST API in the web server \cite{fielding-2000}. For this, the web server translates HTTP requests into MongoDB
queries, performs a query, and returns data to the clients. Alternatively, the client may wish to receive updated data
continuously. To facilitate this the web-server can instantiate a so-called ChangeStream \footnote{\url{https://docs.mongodb.com/manual/changeStreams/}},
which provides a subscription to changes in selected data. That is, the database will automatically forward certain data to
the web server using the ChangeStream. The web server then forwards this data to the user client using websockets \cite{websockets}
so that it receives that data as it is updated. This way, analyzed data can be forwarded from the \emph{processor} to 
a user client as soon as it has been written to the database.

A video that demonstrates the web based visualization is presented in the online supplemental material to this article. The demonstration
shows ECEi data that has been bandpass filtered, with a pass band extending from $5$ to $9$ kHz, such that a rotating magnetic island
becomes visible. The magnetic island segmentation model introduces in \secref{applications} is incorporated in the dashboard and can
be used on-the-fly to mark the different regions of the radial phase inversion structure. This demonstrates how information from trained
machine learning models can be incorporated into data visualization workflows. Future applications of \DDELTA may include a hot-spot
detection, where a machine-learning model would overlay detected hot-spots on camera data streams. Machine operators could use such
information to optimize plasma positioning for the next discharge.

\section{Conclusions and future work}
\label{sec:conclusions}
We have introduced the \DDELTA framework that aims to facilitate near real-time streaming analysis and visualization of big and
fast fusion data. We demonstrated that this framework allows to stream large imaging data sets from remote fusion
experiments to HPC centers, perform computationally demanding data analysis tasks, and visualize data analysis results immediately
after a plasma shot is done and in time before the next one.

\DDELTA is a distributed system, designed using an object oriented architecture and implemented in python. Data analysis is
performed using linear workflow, which is readily extensible with custom pre-processing and data analysis kernels and
leverages distributed HPC resources using a master-worker architecture. By writing data analysis results into a networked
database, the analyzed data can be ingested by visualization dashboard as it is written. Operating \DDELTA alongside
experiments allows to execute computationally demanding turbulence analysis in between shows. By incorporating
machine learning models into the visualization sub-system it also allows to present augmented data visualizations
on web-based dashboards immediately after data from a shot has been analyzed. Focusing on ECEi data from KSTAR, we
show that even a simple U-Net architecture robustly identifies radial phase inversion structures in the images,
which are footprints of magnetic islands.

We have presented extensive benchmarks of the example turbulence workflow. Executing the workflow takes the same amount of time
when the data is streamed from KSTAR to NERSC and when it is loaded from local storage at NERSC. The master-worker architecture
balances compute tasks across the available compute resources of the compute allocation and we find a strong scaling of the walltime
with the amount of compute resources for up to 16 GPUs. For larger resource allocations we find that on-node compute power limits this
range of the scaling.

Future work aims to expand the use-cases for streaming data analytics in fusion sciences. In particular we aim to include
other data sources such as bolometers and Mirnov coils, both of which would motive other data analysis cases. Bolometry
data could for example be used to automate bolometry inversion and possible integrate this data with
on-line training for surrogate machine learning models \cite{carvalho-2019, ferreira-2020}. Another possible route to expand this
work is to focus on running sophisticated integrated modelling tasks in between shots. This approach may require to
utilize more fine-grained resource distribution libraries, such as Ray \cite{liaw-2018}.





\section*{Acknowledgement(s)}
This research used resources of the National Energy Research Scientific Computing Center (NERSC), a U.S. Department of Energy Office of Science User Facility located at Lawrence Berkeley National Laboratory, operated under Contract No. DE-AC02-05CH11231. Delta is available on github \footnote{\url{https://github.com/rkube/delta}}.

\bibliography{myrefs}

\end{document}